\documentclass{elsarticle}

\usepackage{graphicx}

\usepackage{amsmath,amsfonts,amssymb,eucal}

\usepackage{subfigure}

\usepackage[notablist,tablesonly,nomarkers]{endfloat}

\newtheorem{e-proposition}[theorem]{Proposition}

\newtheorem{e-definition}[theorem]{Definition\rm}

\def\og{\leavevmode\raise.3ex\hbox{$\scriptscriptstyle\langle\!\langle$~}}
\def\fg{\leavevmode\raise.3ex\hbox{~$\!\scriptscriptstyle\,\rangle\!\rangle$}}

\begin{document}

\begin{frontmatter}

\title{Transient rolling friction model for discrete element simulations
       of sphere assemblies}

\author{Matthew R. Kuhn}
\ead{kuhn@up.edu}
\address{Dept.\ of Civil Engrg.,
         Donald P.\ Shiley School of Engrg., Univ.\ of Portland,
         5000 N.\ Willamette Blvd.,
         Portland, OR  97203, U.S.A. 
         Tel.~1-503-943-7361. Fax~1-503-943-7316. kuhn@up.edu}


\begin{center}
{\small Received Nov. 14, 2011; accepted after revision March 29, 2013}
\end{center}

\begin{abstract}
The rolling resistance between a pair of contacting particles can
be modeled with two mechanisms.
The first mechanism, already widely
addressed in the DEM literature, involves a contact moment between
the particles.
The second mechanism involves a reduction of the
tangential contact force, but without a contact moment.
This type of rotational resistance, termed creep-friction, is the subject
of the paper.
Within the creep-friction literature, the term ``creep''
does not mean a viscous mechanism, but rather connotes a slight slip
that accompanies rolling.
Two extremes of particle motions bound the
range of creep-friction behaviors: a pure tangential translation is
modeled as a Cattaneo-Mindlin interaction, whereas prolonged steady-state
rolling corresponds to the traditional wheel-rail problem described
by Carter, Poritsky, and others.
DEM simulations, however, are dominated
by the transient creep-friction rolling conditions that lie between
these two extremes.
A simplified model is proposed for the three-dimensional
transient creep-friction rolling of two spheres.
The model is an extension of
the work of Dahlberg and Alfredsson, who studied the two-dimensional
interactions of disks. The proposed model is applied to two different
systems: a pair of spheres and a large dense assembly of spheres.
Although creep-friction can reduce the tangential contact force that would
otherwise be predicted with Cattaneo-Mindlin theory, a significant
force reduction occurs only when the rate of rolling is much greater than
the rate of translational sliding and only after a sustained period
of rolling.
When applied to the deviatoric loading of an assembly
of spheres, the proposed creep-friction model has minimal effect on
macroscopic strength or stiffness. At the micro-scale of individual
contacts, creep-friction does have a modest influence on the incremental
contact behavior, although the aggregate effect on the assembly's
behavior is minimal.
%
\keyword{Contact~mechanics; granular~materials; rolling; Hertz.}
\vskip 0.5\baselineskip

} 

\end{abstract}
\end{frontmatter}

\section{Introduction}

In a Discrete Element (DEM) simulation of a granular material, each
particle is represented as a discrete object that interacts with neighboring
particles at its contacts. 
Each contact interaction yields the contact
force that results from the movements of the two particles. 
Some recent
DEM codes also include a possible \emph{contact moment} that arises
from the pair-wise rotational interactions of the particles. With
this form of \emph{rotational resistance}, the contact interaction
is stiffened between the particle pairs, and the entire assembly is
hardened and strengthened as a result. 
Besides imparting a macro-scale
hardening, 
Iwashita and Oda~\cite{Iwashita:1998a} 
showed that contact
moments also alter the internal length scale of an assembly, an effect
that is evidenced by a change in the thickness and character of localization
features such as shear bands. 
Ai et al.~\cite{Ai:2011a} provide
a comprehensive survey of the many mechanisms that have been proposed
for implementing a contact moment in DEM codes, mechanisms that include
various combinations of rotational springs and rotational dissipation
elements, such as rotational dampers and sliders. 
In the Paper, we
consider an alternative form of rolling resistance and dissipation
that does not involve contact moments. 
This mechanism, termed \emph{creep-friction},
has its origins in the rolling friction literature and results from
micro-slip at particle contacts. Within the creep-friction literature,
the term ``creep'' does not mean a viscous mechanism, but rather connotes
a gradual slip during rolling. 
Creep-friction has, as yet, received
little attention within the DEM community. We begin by clarifying
the distinction between two general forms of rotational resistance:
contact moments and creep-friction. 
We then focus on the creep-friction
mechanism of rotational resistance and describe its similarities and
differences with translational resistance, in particular that of Cattaneo~\cite{Cattaneo:1938a}
and 
Mindlin~\cite{Mindlin:1949a}, 
a type of sliding resistance that
is widely used in DEM codes. 
Although, the creep-friction mechanism
has received much interest over the past century, no exact solutions
have been offered for many basic problems in two-dimensional interaction,
let alone for the general three-dimensional conditions that one encounters
in granular flows. The paper proposes an approximate model for creep-friction
(Section~\ref{sec:model})
and then uses this model to gauge the relative importance of creep-friction
in common granular phenomena (Section~\ref{sec:implementation}).

\subsection{Two categories of rolling resistance}

\begin{figure}
\begin{centering}
\includegraphics{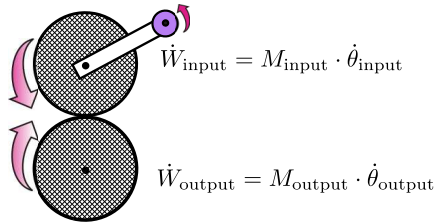}
\par\end{centering}
\caption{Input and output power, with $M_{\text{{input}}}\cdot\dot{\theta}_{\text{input}}\geq M_{\text{{output}}}\cdot\dot{\theta}_{\text{output}}$\label{fig:Input-and-output}}
\end{figure}

Two categories of rolling resistance can be distinguished by considering
a pair of smooth cylindrical rollers of equal radius, one being driven
by the other (Fig.~\ref{fig:Input-and-output}).%
Various mechanisms can lead to energy dissipation, causing the input
power $M_{\text{{input}}}\cdot\dot{\theta}_{\text{input}}$ to exceed
the output power $M_{\text{{output}}}\cdot\dot{\theta}_{\text{output}}$
(here, $M$ and $\dot{\theta}$ represent moments and rotational velocities).
With the first category of rolling resistance, a contact moment acts
between the two rollers so that the input moment exceeds the output
moment, yet preserving the rotational velocities. 
This type of rolling
resistance was used in the DEM simulations of 
Iwashita and Oda~\cite{Iwashita:1998a}
and is now available in many DEM codes 
(see Ai et al.~\cite{Ai:2011a}
for a survey). 
The presence of a contact moment is most obvious in
the interaction of two gears that rotate with synchronous velocities,
but in which the teeth rub in manner that causes a torque reduction
between the two gears: a reduction that can be modeled as a contact
moment. A contact moment is also produced when the two contacting
materials are inelastic, which results in an asymmetry of stress within
the small contact area between the two rolling objects. As rolling
proceeds, the leading portion of the contact is continually loaded
while the trailing portion is unloaded, and any difference in the
loading and unloading stiffnesses (as a result of plasticity, viscosity,
crushing, or sticky adhesion) will produce a small contact moment.
\par
The second category of rolling resistance, 
also illustrated in Fig.~\ref{fig:Input-and-output},
is the focus of the Paper. 
With this mechanism, the absence of a
contact moment preserves the torque (that is,
$|M_{\text{{input}}}|=|M_{\text{{output}}}|$),
but the output rotational velocity is diminished between the driving
roller and the driven roller. 
This latter type of resistance, termed
creep-friction, results from micro-slip between the two rollers within
their small contact region, causing energy dissipation in the absence
of a contact moment. 
Both 
Reynolds~\cite{Reynolds:1876a} 
and 
Johnson~\cite{Johnson:1985a}
likened this type of micro-slip dissipation to the slip that occurs
between a belt and its pulley. 
Creep-friction is the dominant dissipation
mechanism in most rolling stock and, hence, received early attention
in the investigation of rail-wheel interactions. 
As a train engine
rolls steadily upon its rails, the wheel rim moves slightly faster
than the train itself, and this small disparity increases as the grade
steepens until, in the limit, the train stalls as the wheels spin
futilely upon their rails. 
The paper concerns this form of rolling
resistance and dissipation.

\subsection{Creep-friction vs. Cattaneo-Mindlin sliding}

\begin{figure}
\begin{centering}
\includegraphics{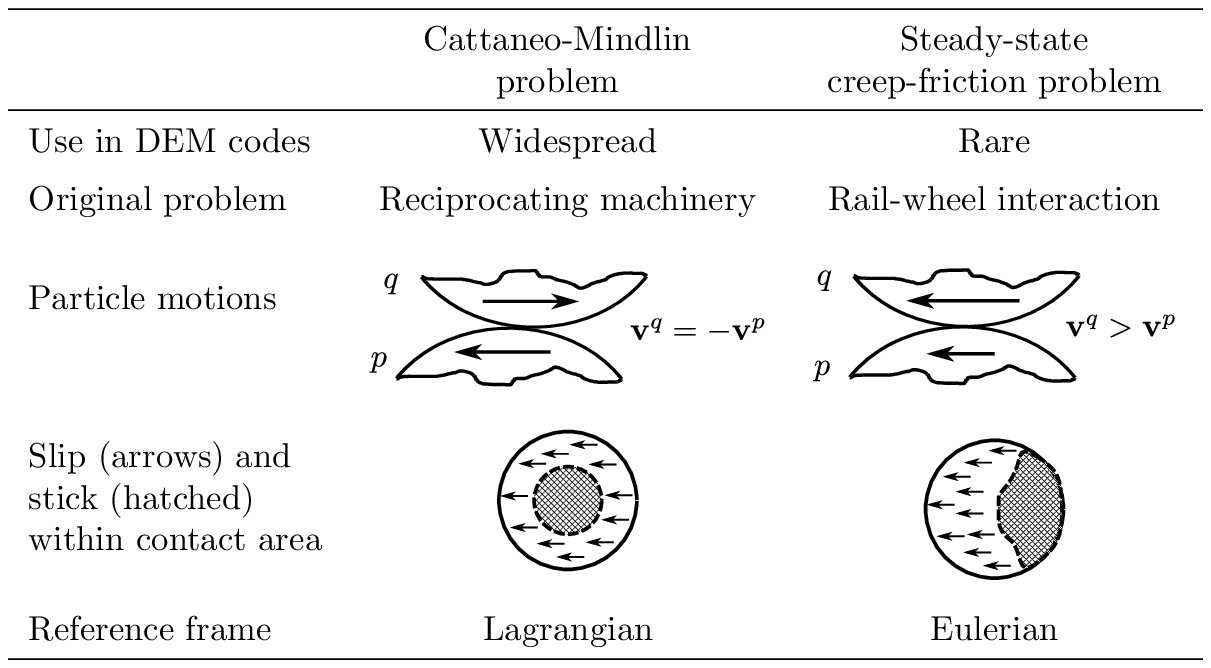}
\par\end{centering}
\caption{Comparison of the Cattaneo-Mindlin and steady-state creep-friction
problems.\label{fig:Comparison}}
\end{figure}

Although both result from micro-slip within the contact zone, a distinction
must be made between the mechanics of creep-friction rolling and that
of translational sliding (see Figure~\ref{fig:Comparison} 
for a summary of these differences).
The problem of translational interaction between two elastic spheres
was independently solved by 
Cattaneo~\cite{Cattaneo:1938a} 
and 
Mindlin~\cite{Mindlin:1949a},
and this purely translational problem will be referred 
to as the \emph{Cattaneo-Mindlin problem}. 
Its solution is in the form of a relationship between the
translational (tangential) displacement and the tangential traction
and force. 
This relationship is known to be complexly path-dependent,
such that the tangential force depends upon the previous sequence
of both tangential and normal displacements. 
Mindlin and Deresiewicz~\cite{Mindlin:1953a}
developed solutions for eleven sequences of such loading and unloading,
and these and other solutions of the translational problem have been
widely incorporated into DEM codes (e.g., those of 
Thornton~\cite{Thornton:1988a};
Lin and Ng~\cite{Lin:1997a}; 
Vu-Quoc and Zhang~\cite{Vuquoc:1999a}; 
and
Kuhn~\cite{Kuhn:2011a}). 
\par
The creep-friction mechanism was first recognized 
by 
Reynolds~\cite{Reynolds:1876a},
but its investigation proceeded independently from that of the translational
Cattaneo-Mindlin problem. 
The two-dimensional problem of creep-friction
between two elastic rolling cylinders was solved by 
Carter~\cite{Carter:1926a}
and 
Poritsky~\cite{Poritsky:1950a} 
in the form of an exact relationship
between the tangential force and the ratio of the rotational velocities
of the two cylinders. 
The three-dimensional problem of rolling between
elastic spheres was later addressed in the thesis of 
Kalker~\cite{Kalker:1967a},
who developed an exact solution in the form of a series of elliptic
functions. 
In subsequent work, Kalker developed solutions for a breadth
of creep-friction problems, including solutions for a wider range
of rotational velocities, solutions for different contact profiles,
solutions for dissimilar bodies, and solutions for the simultaneous
rolling and twisting (spin) of two bodies 
(see \cite{Kalker:1990a,Kalker:2000a}
for compendiums of this work).
\par
Figure~\ref{fig:Comparison} gives a cursory comparison of the Cattaneo-Mindlin
and creep-friction problems and their associated mechanics.%
For both problems, the two contacting bodies are isotropically elastic
but have a frictional limit that applies to tangential traction within
the contact area. We assume that the two bodies are spheres of equal
radius and possess identical 
elastic properties~--- the so-called quasi-identical
condition in the creep-friction literature. 
For this situation, the
two spheres touch within a small circular contact area, and the normal
traction within this area is given by the Hertz solution, which applies
to both the Cattaneo-Mindlin and creep-friction problems. 
In the translational
Cattaneo-Mindlin problem, the two particles, labeled ``$p$'' and
``$q$'', move in opposite directions with material 
velocities $\mathbf{v}^{q}=-\mathbf{v}^{p}$
(Fig.~\ref{fig:Comparison}, middle column). 
With creep-friction,
the two velocities differ, perhaps only slightly, with one larger
than the other, say $|\mathbf{v}^{q}|>|\mathbf{v}^{p}|$. 
For the
traditional problem of rail-wheel rolling interaction, this difference
is very small (the wheel rim will typically move faster than the train
by only a fraction of a percent), but solutions have also been developed
for a wide range of velocity differences 
(see Kalker~\cite{Kalker:2000a}).
A difference in the two velocities produces an asymmetric condition
across the contact area, in the sense that material from each body
will enter the contact along one edge of the contact area and will
exit the contact along the opposite edge. 
This situation produces
an asymmetric slip area within the contact: slip occurs near the exit
(trailing) edge, whereas stick (no slip) occurs near the entering
(leading) edge (see Fig.~\ref{fig:Comparison}, right column). 
The situation is different from the Cattaneo-Mindlin problem, in which
slip occurs within a symmetric annular area that surrounds an inner
circular stick region. Because material continually enters and leaves
the contact zone, the rolling creep-friction problem is usually formulated
within the context of a moving frame using Eulerian kinematics; whereas,
the Cattaneo-Mindlin problem is usually solved within a stationary
frame using Lagrangian kinematics. 
The Eulerian approach is described in the following section.

\section{Model formulation}\label{sec:model}

\subsection{Contact kinematics}

\begin{figure}
\begin{centering}
\includegraphics{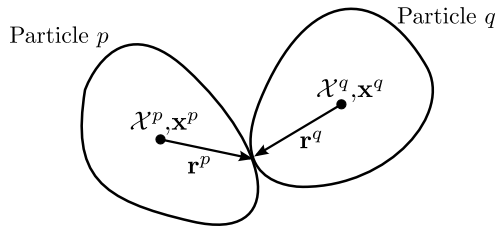}
\par\end{centering}
\caption{Two contacting particles, $p$ and $q$, with incremental movements
$d\mathbf{u}^{p}$, $d\mathbf{u}^{q}$, $d\boldsymbol{\theta}^{p}$,
and $d\boldsymbol{\theta}^{q}$.\label{fig:Two-particles}}
\end{figure}

Material points $\boldsymbol{\mathcal{X}}^{p}$and $\boldsymbol{\mathcal{X}}^{q}$
are at the centers of two smooth particles, $p$ and $q$, and these points
have positions $\mathbf{x}^{p}$ and $\mathbf{x}^{q}$ within the
global reference frame (Fig.~\ref{fig:Two-particles}).%
For the moment, we will assume that the particles are rigid 
and that the contact is an infinitesimal point. 
Vectors $\mathbf{r}^{p}$ and $\mathbf{r}^{q}$ connect the
material points $\boldsymbol{\mathcal{X}}^{p}$and $\boldsymbol{\mathcal{X}}^{q}$
to the contact point. 
Between the times $t$ and $t+dt$, the particles
undergo incremental translation and rotation movements $d\mathbf{u}^{p}$,
$d\mathbf{u}^{q}$, $d\boldsymbol{\theta}^{p}$, and $d\boldsymbol{\theta}^{q}$,
and these four movement vectors are described by their twelve scalar
components, forming a 12-dimensional vector space of possible movements.
Through a suitable transformation of these twelve components, we can
identify six relative movements that are objective, in the sense that
equivalent movements would be reported by two observers having independent
motions 
\cite{Kuhn:2004k,Kuhn:2005c}. 
The remaining six movements
are, in essence, observer-dependent rigid-body motions that are non-objective.
The six objective, relative movements include three translational
movements (which we collect in a ``relative displacement vector,''
$d\mathbf{u}^{\text{disp}}$) and three rotational movements. 
The relative displacement is
\begin{equation}
d\mathbf{u}^{\text{disp}}=d\mathbf{u}^{q}-d\mathbf{u}^{p}+\left(d\boldsymbol{\theta}^{q}\times\mathbf{r}^{q}-d\boldsymbol{\theta}^{p}\times\mathbf{r}^{p}\right)\label{eq:dudisp}
\end{equation}
representing the translation of a point attached to
$q$ at the contact relative to its counterpart on particle $p$,
again assuming that the two particles are rigid.
\par
The three remaining objective movements are associated with relative
rotations. 
Two of these three movements comprise a rolling displacement
vector $d\mathbf{u}^{\text{roll}}$, which represents the average
of the movements of two material points~--- one attached to particle
$p$ near the contact and the other attached to $q$~--- between
the times $t$ and $t+dt$. 
Because the instantaneous motions of two
rigid bodies lie within the tangent plane at their contact, the 3-vector
$d\mathbf{u}^{\text{roll}}$ has only two independent components.
Kuhn and Bagi~\cite{Kuhn:2004b} 
have shown that this rolling vector
is objective and that it can be computed from the incremental motions
of two smooth rigid particles of arbitrary smooth shape, as
\begin{equation}
d\mathbf{u}^{\text{roll}}=\left(\mathbf{K}^{p}+\mathbf{K}^{q}\right)^{-1}\cdot\left[\left(d\boldsymbol{\theta}^{q}-d\boldsymbol{\theta}^{p}\right)\times\mathbf{n}+\frac{1}{2}\left(\mathbf{K}^{p}-\mathbf{K}^{q}\right)\cdot d\overline{\mathbf{u}}^{\text{disp}}\right]\label{eq:durollgeneral}
\end{equation}
where $\mathbf{K}^{p}$ and $\mathbf{K}^{q}$ are the curvature
tensors of the two surfaces; $\mathbf{n}$ is the unit normal vector
of the tangent plane (directed outward from particle $p$); and the
tangential displacement $d\overline{\mathbf{u}}^{\text{disp}}$ is
the projection of $d\mathbf{u}^{\text{disp}}$ onto the tangent plane:
\begin{equation}
du^{\text{n,disp}}=d\overline{\mathbf{u}}^{\text{disp}}\cdot\mathbf{n}\label{eq:dudispnorm}
\end{equation}
\begin{equation}
d\overline{\mathbf{u}}^{\text{disp}}=d\mathbf{u}^{\text{disp}}-du^{\text{n,disp}}\mathbf{n}\label{eq:dudisptan}
\end{equation}
in which $du^{\text{n,disp}}$ is the normal displacement.
We adopt a direction of $d\mathbf{u}^{\text{roll}}$ that is consistent
with the creep-rolling literature: the direction of material passing
through the contact area. 
Equation~(\ref{eq:durollgeneral}) is greatly
simplified when the two surfaces are spherical at the contact:
\begin{equation}
d\mathbf{u}^{\text{roll}}=-\frac{R^{p}R^{q}}{R^{p}+R^{q}}\left(d\boldsymbol{\theta}^{q}-d\boldsymbol{\theta}^{p}\right)\times\mathbf{n}-\frac{1}{2}\frac{R^{p}-R^{q}}{R^{p}+R^{q}}d\overline{\mathbf{u}}^{\text{disp}}\label{eq:durollspheres}
\end{equation}
where $R^{p}$ and $R^{q}$ are the convex surface radii
of the two particles at the contact. Both $d\overline{\mathbf{u}}^{\text{disp}}$
and $d\mathbf{u}^{\text{roll}}$ are 3-vectors that lie in the tangent
plane, each with two independent components.
\par
The third rotational movement is a twisting (torsion, spin) of one
particle relative to the other, taken about the normal vector $\mathbf{n}$.
Although such twisting can alter the distribution of traction within
the contact area (and, hence, has received considerable attention
within the creep-friction literature), it will not be considered in
the Paper, which is confined to the effects of contact displacement
and rolling, $d\mathbf{\overline{\mathbf{u}}}^{\text{disp}}$ and
$d\mathbf{u}^{\text{roll}}$.
\par
The kinematics, thus far, have been limited to rigid particles. 
If the particles were truly rigid, the contact would be a single point;
the relative displacement $d\mathbf{u}^{\text{disp}}$ would not include
any normal component that would violate the particles' impenetrability
(a constraint $du^{\text{n,disp}}=0$); and any tangential component
$d\mathbf{\overline{\mathbf{u}}}^{\text{disp}}$ would produce full
frictional sliding between the particles. 
We now remove the rigidity constraint and permit the particles to deform. 
These particle deformations
will enable contact across a finite area; will give rise to a Hertzian
normal traction within this contact area; and will inhibit tangential
slip, allowing the particles to stick within portions of the contact
area. 
The tangential displacement fields produced by the two particles'
deformations are designated $d\overline{\mathbf{u}}^{p,\text{def}}$
and $d\overline{\mathbf{u}}^{q,\text{def}}$, and these displacement
fields will vary across the contact area between the two particles.
The difference
\[
d\overline{\mathbf{u}}^{\text{def}}=d\overline{\mathbf{u}}^{q,\text{def}}-d\overline{\mathbf{u}}^{p,\text{def}}
\]
will tend to reduce the slip that otherwise would be produced
by the rigid translational displacement $d\overline{\mathbf{u}}^{\text{disp}}$.
Indeed, no slip will occur within those portions of the contact area
where the vector field $d\overline{\mathbf{u}}^{\text{def}}$ fully
counteracts the tangential displacement vector $d\overline{\mathbf{u}}^{\text{disp}}$. 
\par
In general, as the particles move and roll, the contact area will
move across the surfaces of the two particles at the rate $d\mathbf{u}^{\text{roll}}/dt$.
By adopting an Eulerian frame that moves with the contact, the vector
field of tangential slip within the contact area, $d\mathbf{s}$, is
\begin{equation}
d\mathbf{s}=d\overline{\mathbf{u}}^{\text{disp}}-\frac{\partial\overline{\mathbf{u}}^{\text{def}}}{\partial t}dt-\left(\frac{\partial\mathbf{u}^{q}}{\partial\mathbf{x}}-\frac{\partial\mathbf{u}^{p}}{\partial\mathbf{x}}\right)\cdot d\mathbf{u}^{\text{roll}}\label{eq:convection}
\end{equation}
For the non-sliding (stick) portions of the contact area,
$d\mathbf{s}=\mathbf{0}$, but $d\mathbf{s}$ is non-zero elsewhere
within the slip portion of the contact area. 
The final term in Eq.~(\ref{eq:convection})
represents a ``convection of tangential displacement'' into the contact
area, as material is moved (in effect, rolled) through the area. 
The gradient tensors $\partial\mathbf{u}/\partial\mathbf{x}$ represent
planar deformations (stretching and shearing) of the two particles
within the contact plane. Equation~(\ref{eq:convection}) is widely
used within the creep-friction literature, although usually in a scalar
form 
\cite{Johnson:1985a,Kalker:1990a,Kalker:2000a}. 
In keeping with
convention, the displacement $d\overline{\mathbf{u}}^{\text{def}}$
is written as a derivative $\partial\:/\partial t$ within the moving
frame to emphasize its transient nature, as will be discussed below.
Although $d\mathbf{s}$ and $d\overline{\mathbf{u}}^{\text{def}}$
are vector fields across the contact region 
and the $\partial\mathbf{u}/\partial\mathbf{x}$
are tensor fields across the contact region, both $d\overline{\mathbf{u}}^{\text{disp}}$
and $d\mathbf{u}^{\text{roll}}$ are merely 3-vectors that result
from the incremental particle motions. 
These vectors can be readily
computed within each time step of a DEM algorithm, 
with Eqs.~(\ref{eq:dudisp})--(\ref{eq:durollspheres}).

\subsection{Creep-friction and Cattaneo-Mindlin solutions}

The fundamental aim of elastic contact mechanics is finding the deformations,
stresses, and tractions as solutions to a problem of three-dimensional
elasticity. 
In this problem, Eq.~(\ref{eq:convection}) serves as
a displacement boundary condition across the surface of a single particle
body within its contact area. 
The remaining boundary conditions include
a requirement of zero traction outside of the contact area; a requirement
that the tangential traction does not exceed the friction limit within
those non-sliding portions of the contact area 
(wherever $d\mathbf{s}=\mathbf{0}$);
a requirement that for those portions in which the friction limit
is exceed, the direction of $d\mathbf{s}$ is aligned with the tangential
traction; and a requirement that the normal displacements of the two
particles comply within the contact area. 
\par
In the Cattaneo-Mindlin problem, two particles are assumed to move
in opposite directions at their contact, 
with $d\mathbf{u}^{\text{roll}}=\mathbf{0}$,
thus leaving only the first and second terms on the right of the boundary
condition~(\ref{eq:convection}) 
(see the middle column of Fig.~\ref{fig:Comparison}).
On the other hand, the classic problem in creep-friction assumes that
rolling has attained a \emph{steady-state condition} in which the
second term vanishes, leaving only the first and third terms on the
right of Eq.~(\ref{eq:convection}). 
The steady-state condition would apply to a train moving at
constant speed: to an observer who moves with the rail-wheel contact,
the rail and wheel appear to pass through a stationary contact zone,
and deformations within the rail and wheel appear to be constant within
the moving frame, 
with $\partial\overline{\mathbf{u}}^{\text{def}}/\partial t=\mathbf{0}$.
The situation in DEM simulations is far more complex than either
of these extreme cases, since particles will both roll and slide in
bewildering sequences that produce loading and unloading in both normal
and tangential directions, all within a three-dimensional setting.
The term \emph{transient conditions} is used within the creep-friction
literature for cases in which all three terms are active within the
boundary condition given by Eq.~(\ref{eq:convection}). 
Only the
simplest of these transient conditions have been addressed in the
literature.
\par
In a later section, an approximate solution of the transient problem
will be proposed for the contact of two spheres. 
This approximation
is an amalgam of separate solutions of the Cattaneo-Mindlin problem
and of the steady-state creep-friction problem. 
As for the former,
the Cattaneo-Mindlin problem has been largely solved in the works
of Cattaneo, Mindlin, Deresiewicz, and others 
\cite{Cattaneo:1938a,Mindlin:1949a,Mindlin:1953a,Vuquoc:1999a,Jager:2005a,Kuhn:2011a},
and these solutions have been widely incorporated in DEM codes. 
Omitting the details, 
solutions of the Cattaneo-Mindlin problem can be represented as
\begin{equation}
d\overline{\mathbf{u}}^{\text{def}}=\mathcal{C}_{\text{C-M}}\left(d\mathbf{Q}\right)\quad\text{or}\quad d\mathbf{Q}=\mathcal{C}_{\text{C-M}}^{-1}\left(d\overline{\mathbf{u}}^{\text{def}}\right)\label{eq:dudef}
\end{equation}
in which $d\mathbf{Q}$ is the 3-vector incremental change
in the tangential contact force that corresponds to a given tangential
displacement increment $d\overline{\mathbf{u}}^{\text{disp}}$ (in
conventional DEM codes, $d\overline{\mathbf{u}}^{\text{disp}}$ and
$d\overline{\mathbf{u}}^{\text{def}}$ are assumed equal, 
an assumption that does not apply to transient creep-friction).
The symbol $\mathcal{C}_{\text{C-M}}$
represents a tangential compliance function based upon Cattaneo-Mindlin
(C-M) theory, and its inverse $\mathcal{C}_{\text{C-M}}^{-1}$ is
the corresponding tangential contact stiffness function. 
In the context
of a DEM code, $\mathcal{C}_{\text{C-M}}^{-1}$ represents an algorithm
(function, subroutine, procedure, etc.) that computes an output $d\mathbf{Q}$
from an input $d\overline{\mathbf{u}}^{\text{disp}}$. 
This compliance
function will depend upon the elastic coefficients, the friction coefficient,
the normal force, the concurrent normal displacement ($=du^{\text{n,disp}}$),
and the contact's loading history.
\par
As for the steady-state creep-friction problem, approximate solutions
for elastic spheres were developed in the 1960's by Haines and Ollerton~\cite{Haines:1963a}
and by 
Vermeulen and Johnson~\cite{Vermeulen:1964a}. 
Although not
a closed-form solution, 
Kalker~\cite{Kalker:1967a} 
derived an exact
solution in the form of a series of elliptic functions and also developed
software for implementing this solution 
\cite{Kalker:1982a}. 
With the steady-state problem, rolling proceeds steadily in one direction,
and the various creep-friction solutions all have the general scalar
form
\begin{equation}
\xi_{\text{s-s}}\equiv\frac{v^{\text{disp}}}{v^{\text{roll}}}=\mathcal{F_{\text{s-s}}}\left(\frac{|\mathbf{Q}|}{\mu P}\right)\label{eq:GeneralCreepage}
\end{equation}
where $\xi_{\text{s-s}}$ is termed the \emph{steady-state
creepage} and $\mathcal{F_{\text{s-s}}}$ is the corresponding creepage
function. 
During steady-state creep-friction, the velocity $v^{\text{disp}}$
(or the corresponding increment $d\overline{\mathbf{u}}^{\text{disp}}$)
represents a continual slipping of the contact as it rolls at rate
$v^{\text{roll}}$ 
(or with the corresponding increment $d\mathbf{u}^{\text{roll}}$).
For the traditional wheel-rail problem, the ratio $\xi_{\text{s-s}}$
of the displacement (slipping) and rolling velocities is usually small,
often less than one percent.
The creepage function $\mathcal{F_{\text{s-s}}}$ depends upon the
elastic coefficients, the normal force $P$, the friction coefficient
$\mu$, and the ratio of the tangential force to the limiting force
(the product of $\mu$ and $P$). 
The creepage function $\mathcal{F}_{\text{s-s}}$
increases monotonically with the ratio $|\mathbf{Q}/\mu P|$. As in
the previous analogy, the tractive force $\mathbf{Q}$ will increase
when a locomotive engine moves up a steepening grade, causing the
micro-slip creepage $\xi_{\text{s-s}}$ to increase as well.
\par
Both 
Carter~\cite{Carter:1926a} 
and 
Poritsky~\cite{Poritsky:1950a}
developed exact solutions for the creepage function $\mathcal{F}_{\text{s-s}}$
of the two-dimensional cylinder-cylinder problem, which has served
as a basis for wheel-rail rolling. 
Kalker~\cite{Kalker:2000a} 
developed
the following approximation of the creepage function for the three-dimensional
sphere-sphere problem:
\begin{equation}
\mathcal{F_{\text{s-s}}^{\text{Kalker}}}\left(\frac{|\mathbf{Q}|}{\mu P}\right)=\frac{3\mu P}{Ga^{2}C_{11}}\left[1-\left(1-\frac{|\mathbf{Q}|}{\mu P}\right)^{1/3}\right]\label{eq:KalkerCreepage}
\end{equation}
in which $G$ is the shear modulus, $a$ is the radius of
the contact area, and $C_{11}$ is a coefficient that depends upon
the Poisson ratio $\nu$ (Table~\ref{table:C11}). 
\begin{table}
\caption{Kalker's coefficient $C_{11}$ in Eqs.~(\ref{eq:KalkerCreepage})
and~(\ref{eq:KalkerCreepage2}) as a function of Poisson ratio $\nu$
for sphere-sphere contacts
\protect\cite{Kalker:1967a}.}
\label{table:C11}
\centering
\begin{tabular}{cc}
\hline 
Poison ratio & Kalker coefficient\\
$\nu$ & $C_{11}$\\
\hline 
0 & 3.40\\
0.25 & 4.12\\
0.50 & 5.20\\
\hline 
\end{tabular}
\end{table}
Kalker's creepage function was shown to be exact for small creepage
rates $\xi_{\text{s-s}}$ (a situation termed \emph{linear creep},
corresponding to small values of $|\mathbf{Q}|/\mu P$), and it also
correctly predicts gross sliding at the friction limit 
(that is, $\xi_{\text{s-s}}\rightarrow\infty,$
as $|\mathbf{Q}|\rightarrow\mu P$). 
For intermediate values of $|\mathbf{Q}|/\mu P$,
Kalker~\cite{Kalker:1982a} 
showed that Eq.~\ref{eq:KalkerCreepage}
closely fits the experimental data of 
Johnson~\cite{Johnson:1985a}.
\par
With Hertzian contact, the $P$, $G$, and $a$ are inter-related,
as $P=8Ga^{3}/(3R(1-\nu))$, and Kalker's Eq.~(\ref{eq:KalkerCreepage})
can be expressed in the alternative form,
\begin{equation}
\mathcal{F_{\text{s-s}}^{\text{Kalker}}}\left(\frac{|\mathbf{Q}|}{\mu P}\right)=\frac{8\mu}{(1-\nu)C_{11}}\frac{a}{R}\left[1-\left(1-\frac{|\mathbf{Q}|}{\mu P}\right)^{1/3}\right]\label{eq:KalkerCreepage2}
\end{equation}
where $R$ is the particle radius and $\nu$ is the Poisson
ratio. 
This form is instructive, as it shows that the creepage is
non-dimensional and depends directly upon the ratio of the contact
and particle radii, $a/R$, a matter that will be discussed in the
Section~{sec:implementation}.

\subsection{Proposed contact model}

Although exact solutions have been developed for the two extremes
of pure Cattaneo-Mindlin sliding and pure steady-state creep-friction,
the situation is far less accomplished for the intermediate, transient
conditions that are likely to predominate in DEM simulations of granular
flow. 
Kalker~\cite{Kalker:1971a,Kalker1971b} 
developed exact solutions
for several transient cases involving cylinders. 
In one case, two
stationary cylinders, initially sustaining a pure translational force
$Q$, then begin to roll while maintaining constant $Q$ and $P$
(appropriately, the sequence is termed ``Cattaneo-to-Carter'').
Kalker found that the transition in traction from pure Cattaneo-Mindlin
sliding to pure steady-state creep friction was complete when the
two cylinders had rolled a distance $2a$, twice the radius of the
contact area. 
Kalker~\cite{Kalker:1973a} 
later developed an approximate
approach for analyzing transient conditions, modeling the contact
area as a Winkler bed of tangential springs (the so-called ``brush
model''). 
This approach was used by 
Al-Bender and De Moerlooze~\cite{AlBender:2008a}
to analyze the case of two spheres, initially pressed together with
normal force $P$ but zero tangential force, that are then rolled,
with each particle rotating at a constant rate, one particle slightly
faster than the other, to maintain a constant creepage ratio $\xi=d\overline{u}^{\text{disp}}/du^{\text{roll}}$.
Full slip was found to occur when the accumulated tangential displacement
$d\overline{u}^{\text{disp}}$ equaled the radius of the contact area,
$a$. 
\par
In a recent paper, 
Dahlberg and Alfredsson~\cite{Dahlberg:2009a}
further simplified the brush approach for the transient rolling of
two cylinders. 
They reformulated Eq.~(\ref{eq:convection}) by setting
the slip $ds=0$, as would apply within the stick portion of the contact
area, and by approximating the remaining terms as
\[
0=d\overline{u}^{\text{disp}}-\frac{1}{k}dQ-\mathcal{F_{\text{s-s}}^{\text{Carter}}}\left(\frac{|Q|}{\mu P}\right)\cdot du^{\text{roll}}
\]
In this scalar form, 
the deformation field $d\overline{\mathbf{u}}^{\text{def}}$
in the second term of Eq.~(\ref{eq:convection}) is approximated
with a scalar version of Eq.~(\ref{eq:dudef}) using a linear compliance
$1/k$ for the function $\mathcal{C}_{\text{C-M}}$. 
The final, convection
term in Eq.~(\ref{eq:convection}) is approximated as the steady-state
creepage function $\mathcal{F_{\text{s-s}}^{\text{Carter}}}$ multiplied
by the rolling increment 
$du^{\text{roll}}$ (the exact 
Carter~\cite{Carter:1926a}
creep function was used in this two-dimensional analysis of cylinder
rolling). 
With these approximations, the increment of tangential force
$dQ$ is computed as
\begin{equation}
dQ=k\left[1-\mathcal{F_{\text{s-s}}^{\text{Carter}}}\left(\frac{|Q|}{\mu P}\right)\frac{du^{\text{roll}}}{d\overline{u}^{\text{disp}}}\right]d\overline{u}^{\text{disp}}\label{eq:Dahlberg}
\end{equation}
This form is well suited for displacement driven computations,
such as those in DEM codes, in which the contact force must be calculated
from the rolling and sliding displacements within each time step. 
\par
\begin{figure}
\begin{centering}
\includegraphics[scale=0.18]{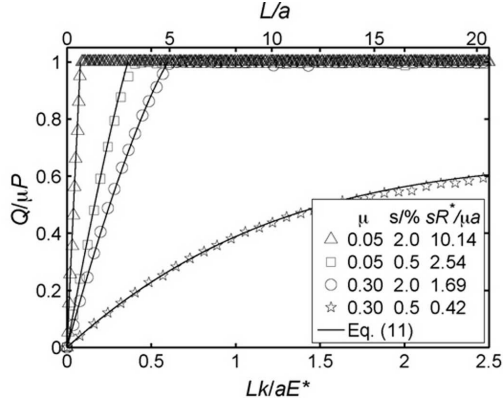}
\par\end{centering}
\caption{Results of 
Dahlberg and Alfredsson~\protect\cite{Dahlberg:2009a} 
for the
rolling of two cylinders. The cylinders are initially pressed together
with force $P$ and are then rolled with a constant creepage ratio
$\xi$: here, $s$ is the equivalent of $\xi$, $L$ is the cumulative
rolled distance, and $E^{\ast}$is $E/2(1-\nu^{2})$. Other symbols
are from the approximate model of Eq.~(\ref{eq:Dahlberg}), and the
lines are the FEM solutions of Dahlberg and Alfredsson.\label{fig:Dahlberg}}
\end{figure}

Dahlberg and Alfredsson~\cite{Dahlberg:2009a} 
validated their approximation
by comparing it with results of a finite element (FEM) model of two
identical elastic cylinders. 
Besides confirming that their model closely
approximates Kalker's Cattaneo-to-Carter solution, they also modeled
the problem of two cylinders that are initially pressed together with
force $P$ and are then rolled with a constant creepage ratio $\xi=d\overline{u}^{\text{disp}}/du^{\text{roll}}$.
Figure~\ref{fig:Dahlberg}%
is adopted from their paper and shows the results from both Eq.~(\ref{eq:Dahlberg})
and from their FEM model. 
The cumulative rolling distance 
$L=\int du^{\text{roll}}=(1/\xi)\int d\overline{u}^{\text{disp}}$
is normalized by dividing by the contact radius, $a$. The results
are for two creepage ratios $\xi$ and for two friction coefficients
$\mu$, as shown. 
For large creepage ratios and small friction coefficients,
nearly full slip develops (with tangential force $Q$ close to $\mu P$)
at rolling distances of less than $5a$. 
With a small creepage ratio
and large friction coefficient, the eventual $Q$ approaches a reduced
condition that applies to steady-state creep-friction, but only after
undergoing large rolling displacements.
\par
An approximate three-dimensional contact model is now developed by
adapting the approach of Dahlberg and Alfredsson to sphere-sphere
contacts. 
The first term on the right of Eq.~(\ref{eq:convection}),
$d\overline{\mathbf{u}}^{\text{disp}}$, is the 3-vector of rigid-body
tangential displacement, as computed with Eqs.~(\ref{eq:dudisp})
and~(\ref{eq:dudisptan}). 
We approximate the second term on the
right of Eq.~(\ref{eq:convection}) as the tangential deformation
that would result from a force increment $d\mathbf{Q}$ if the contact
were undergoing a pure sliding motion: the deformation $d\overline{\mathbf{u}}^{\text{def}}$
computed with the Cattaneo-Mindlin compliance function 
$\mathcal{C}_{\text{C-M}}$
(shown symbolically in Eq.~\ref{eq:dudef}). 
The final term in Eq.~(\ref{eq:convection})
is approximated as the creepage displacement 
$\mathcal{F_{\text{s-s}}}du^{\text{roll}}$
that would result from a rolling increment $du^{\text{roll}}$ if
the contact were undergoing pure steady-state creep-friction. 
In a three-dimensional setting, 
the scalar creepage $\mathcal{F_{\text{s-s}}}du^{\text{roll}}$
is assumed to take the tensor form
\begin{equation}
\mathcal{F_{\text{s-s}}}du^{\text{roll}}\rightarrow\left(\mathcal{F}_{\text{s-s}}\mathbf{q}\otimes\mathbf{p}\right)\cdot d\mathbf{u}^{\text{roll}}\label{eq:Fssduroll}
\end{equation}
such that
\begin{equation}
\left(\frac{\partial\mathbf{u}^{q}}{\partial\mathbf{x}}-\frac{\partial\mathbf{u}^{q}}{\partial\mathbf{x}}\right)\cdot d\mathbf{u}^{\text{roll}}\approx\left(\mathcal{F}_{\text{s-s}}\mathbf{q}\otimes\mathbf{p}\right)\cdot d\mathbf{u}^{\text{roll}}\label{eq:ApproxConvection}
\end{equation}
where $d\mathbf{u}^{\text{roll}}$ is the incremental rolling
vector computed with Eq.~(\ref{eq:durollgeneral}) or~(\ref{eq:durollspheres}).
The creep function $\mathcal{F_{\text{s-s}}}$ gives the scalar magnitude
of creep displacement, whereas the dyad $\mathbf{q}\otimes\mathbf{p}$
carries information about the unit direction $\mathbf{q}$ of the
creep displacement and the unit direction $\mathbf{p}$ upon which
$d\mathbf{u}^{\text{roll}}$ is projected. 
That is, 
tensor $\left(\mathcal{F}_{\text{s-s}}\mathbf{q}\otimes\mathbf{p}\right)$
is a rank-1 linear transformation from 
the rolling increment $d\mathbf{u}^{\text{roll}}$
to the creep displacement. 
We further assume that creepage proceeds
in the direction of the shearing force $\mathbf{Q}$, such that
\begin{equation}
\mathbf{p}=\frac{\mathbf{Q}}{|\mathbf{Q}|}\label{eq:p}
\end{equation}
and that the creepage displacement is prodded by the full
magnitude of the rolling increment, such that
\begin{equation}
\mathbf{q}=
\frac{d\mathbf{u}^{\text{roll}}}{|d\mathbf{u}^{\text{roll}}|}\label{eq:q}
\end{equation}
Following the approach of Dahlberg and Alfredsson, we set
$d\mathbf{s}=\mathbf{0}$ in Eq.~(\ref{eq:convection}) and substitute
Eqs.~(\ref{eq:dudef}) and (\ref{eq:Fssduroll})--(\ref{eq:q}),
with the resulting approximation
\begin{align}
d\mathbf{Q} & =\mathcal{C}_{\text{C-M}}^{-1}\left(d\overline{\mathbf{u}}^{\text{disp}}-\mathcal{F_{\text{s-s}}^{\text{Kalker}}}\left(\frac{|\mathbf{Q}|}{\mu P}\right)\,\frac{\mathbf{Q}}{|\mathbf{Q}|}\,|d\mathbf{u}^{\text{roll}}|\right)\nonumber \\
 & =\mathcal{C}_{\text{C-M}}^{-1}
\left(d\overline{\mathbf{u}}^{\text{disp}}
 -\frac{8\mu}{(1-\nu)C_{11}}\frac{a}{R}
 \left(1-\left(1-\frac{|\mathbf{Q}|}{\mu P}\right)^{1/3}\right)\,
 \frac{\mathbf{Q}}{|\mathbf{Q}|}\,|d\mathbf{u}^{\text{roll}}|\right)
\label{eq:dQ3Db}
\end{align}
where Kalker's creepage function, 
$\mathcal{F_{\text{s-s}}^{\text{Kalker}}}(|\mathbf{Q}|/(\mu P))$,
is used for the rolling of two identical spheres 
(see Eqs.~\ref{eq:KalkerCreepage}
and~\ref{eq:KalkerCreepage2}).
\par
Although Eq.~(\ref{eq:dQ3Db}) is merely an approximate solution
of a complex three-dimensional boundary-value problem, it does match
exact solutions in two respects. 
After an extended absence of rolling
$d\mathbf{u}^{\text{roll}}$, the equation gives the exact Cattaneo-Mindlin
result for sliding displacements $d\overline{\mathbf{u}}^{\text{disp}}$
(Eq.~\ref{eq:dudef}). 
Moreover, after an extended period of constant creepage,
$\xi=|d\overline{\mathbf{u}}^{\text{disp}}|/|d\mathbf{u}^{\text{roll}}|$,
the equation gives the exact steady-state creep-friction result 
(Eqs.~\ref{eq:GeneralCreepage}
and~\ref{eq:KalkerCreepage2}). 
That is, Eq.~(\ref{eq:dQ3Db}) fits
the two extremes of pure Cattaneo-Mindlin sliding and pure steady-state
creep-friction. 
\par
The creep friction model of Eq.~(\ref{eq:dQ3Db}) differs from contact
models that incorporate rotational springs and sliders. Whereas rotational
springs will stiffen the contact behavior, creep-friction softens
the contact resistance, as a rolling term is subtracted from the translational
displacement $d\overline{\mathbf{u}}^{\text{disp}}$, reducing the
force increment $d\mathbf{Q}$. 
Although this softening might seem
a type of viscosity, the effect of creep-friction depends upon a dimensionless
\emph{ratio} of two objective rates~--- the displacement and rolling
rates~--- as is apparent in 
Eqs.~(\ref{eq:GeneralCreepage}) and~(\ref{eq:dQ3Db}).
This creep-friction model is, therefore, rate-independent and objective.
\par
Equation~(\ref{eq:dQ3Db}) also suggests the relative importance
of translational and rolling movements in affecting the contact force
$\mathbf{Q}$. 
Unlike $d\overline{\mathbf{u}}^{\text{disp}}$, the
rolling movement $d\mathbf{u}^{\text{roll}}$ is multiplied by the
ratio of the contact and particle radii, $a/R$, which will usually
be much smaller than 1. 
For example, with a quartz sand or other geologic
material confined at 1~atmosphere of pressure, the ratio $a/R$ is
less than 2\%, and the tangential force will be influenced much less
by the rolling velocity than by the sliding velocity. 
The effect of
rolling will be much larger, of course, with softer materials confined
at larger pressures.
\par
When used in DEM simulations, the Eq.~(\ref{eq:dQ3Db}) has a number
of attractive features:
\begin{itemize}
\item 
The equation is displacement-driven: it requires the sliding and rolling
increments, $d\overline{\mathbf{u}}^{\text{disp}}$ 
and $d\mathbf{u}^{\text{roll}}$,
to compute $d\mathbf{Q}$. 
DEM codes are displacement-driven and must first
compute $d\overline{\mathbf{u}}^{\text{disp}}$ in order in order
to find the tangent force. 
Equation~(\ref{eq:dQ3Db}) only requires
the added calculation of $d\mathbf{u}^{\text{roll}}$ using Eq.~(\ref{eq:durollgeneral})
or~(\ref{eq:durollspheres}).
\item 
The equation models arbitrary combinations of sliding and rolling.
Moreover, it allows an arbitrary concurrent normal 
displacement $du^{\text{n, disp}}$
and normal force $N$. 
The three increments~--- $d\overline{\mathbf{u}}^{\text{disp}}$,
$d\mathbf{u}^{\text{roll}}$, and $du^{\text{n, disp}}$~--- can
be expected to occur in almost any combination and in almost any sequence
during a DEM simulation.
\item 
The stiffness function $\mathcal{C}_{\text{C-M}}^{-1}$ would already
exist as a procedure or subroutine within an existing DEM code, receiving
the full tangential displacement $d\overline{\mathbf{u}}^{\text{disp}}$
as input and returning $d\mathbf{Q}$ as its output. Eq.~(\ref{eq:dQ3Db})
can utilize an existing code for $\mathcal{C}_{\text{C-M}}^{-1}$
and only requires an amendment of the input $d\overline{\mathbf{u}}^{\text{disp}}$.
\end{itemize}
Vu-Quoc and Zhang~\cite{Vuquoc:1999a} 
proposed a model for rolling
contact that has some similarities with Eq.~(\ref{eq:dQ3Db}). 
Using a different approach, they arrived at a scalar equation for an effective
displacement $\Delta\delta^{\prime}$ that would be used to compute
the force increment $dQ$:
\[
\Delta\delta^{\prime}=d\overline{u}^{\text{disp}}-\frac{2}{\pi}\frac{du^{\text{roll}}}{a}\overline{u}^{\text{disp}}
\]
where $\overline{u}^{\text{disp}}$ is the cumulative tangential
displacement. The proposed equation~(\ref{eq:dQ3Db}) has the advantage
of conforming with established theory when two spheres undergo pure
steady-state creep-friction rolling.

\section{Implementation}\label{sec:implementation}

In this section, we investigate creep-friction and rolling in two
granular systems: (1) a two-particle system that undergoes a simple
combination of sliding and rolling, and (2) a system of 4096 densely
packed spheres loaded in triaxial compression. The purpose is to establish
the relative importance of creep-friction in the behavior of the two
systems.

\subsection{Two-particle system}

\begin{figure}
\centering
\subfigure[]{\begin{raggedright}
             \includegraphics[clip,scale=0.9]{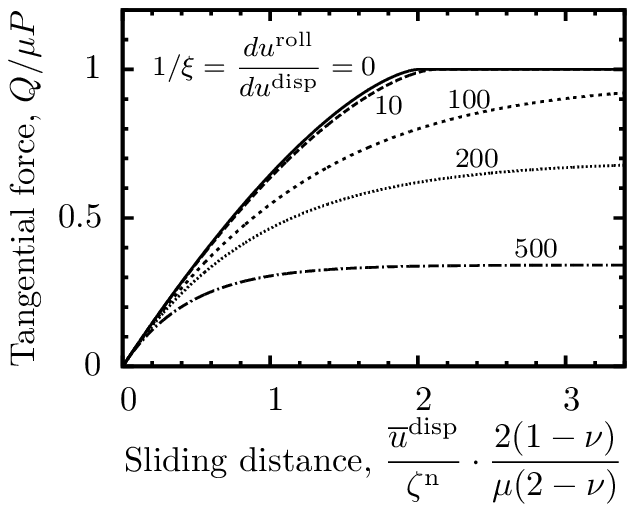}
             \par\end{raggedright}}%
\quad
\subfigure[]{\begin{raggedleft}
             \includegraphics[clip,scale=0.9]{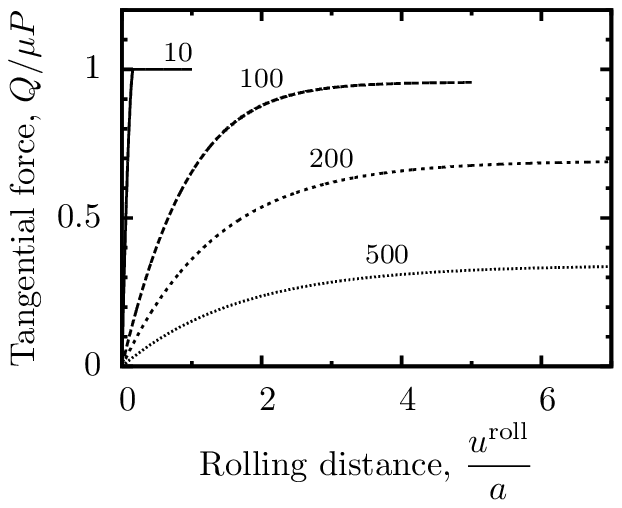}
             \par\end{raggedleft}}
\caption{Results of two spheres that simultaneously slide and roll. The spheres
are initially pressed together with force $P$ and indentation $\zeta^{\text{n}}$
and are then rotated in opposite directions to produce both sliding
$d\overline{u}^{\text{disp}}$ and rolling $du^{\text{roll}}$. Tangential
force is plotted in two forms: (a) with respect to the cumulative
displacement $\overline{u}^{\text{disp}}$ and (b) with respect to
the cumulative rolling distance $u^{\text{roll}}$. Pure sliding corresponds
to $1/\xi\rightarrow0$; pure rolling corresponds to $1/\xi\rightarrow\infty$.\label{fig:2_particles}}
\end{figure}

In the first system, two equal-radius spheres are pressed together
so that the indentation of each particle, $\zeta^{\text{n}}$, is
0.01\% of its radius $R$ (in DEM parlance, the indentation $\zeta^{\text{n}}$
is half of the overlap). 
In this initial state, no tangential force
is applied, and the contact radius ``$a$'' is 1\% of $R$ (with Hertz
contact, the contact radius equals $\sqrt{\zeta^{\text{n}}R}$). 
While maintaining a constant indentation and a constant normal force $N$,
the two spheres are then rotated in opposite directions, one faster
than the other, creating both rolling and tangential displacements
in small increments $du^{\text{roll}}$ and $d\overline{u}^{\text{disp}}$.
A constant creep ratio $\xi=d\overline{u}^{\text{disp}}/du^{\text{roll}}$
is maintained during this process. 
The progressively increasing tangential
displacement causes the tangential force to increase from its initial,
zero condition to an eventual value that is equal to or less than
the friction limit $\mu P$. 
Figure~\ref{fig:2_particles} shows
the advancing tangential force for several different inverse creep
ratios $1/\xi$. 
The tangential force $T$ has been normalized with respect to the
friction threshold $\mu P$. Figure~\ref{fig:2_particles}a shows
the increasing tangential force versus the cumulative displacement
$\overline{u}^{\text{disp}}$ ($=\int d\overline{u}^{\text{disp}}$),
where the displacement has been normalized with respect to the indentation
$\zeta^{\text{n}}$. Fig.~\ref{fig:2_particles}b 
also shows the same tangential
force, but plotted against the cumulative rolling distance $u^{\text{roll}}$,
where the rolling distance has been normalized with respect to the
contact radius ``$a$''. The two different ways of normalizing displacement
are consistent with the two limits of contact behavior. In the extreme
of sliding with no rolling ($1/\xi=0$), the tangential stiffness
only depends upon the normal indentation and the parameters $\mu$
and $\nu$ (see \S2 of 
\cite{Mindlin:1953a}). 
At the other extreme,
predominated by rolling ($1/\xi\rightarrow\infty$), the steady-state
tangential force depends upon the contact radius ``$a$'' (as in Eq.~\ref{eq:KalkerCreepage2}).
\par
Figure~\ref{fig:2_particles}a shows that the rolling rate has negligible
effect upon the tangential force unless the rolling rate is much larger
than the displacement rate, with an inverse ratio $1/\xi$ larger
than~10. 
That is, the eventual tangential force can be greatly reduced
from the friction limit $\mu P$, but a significant reduction will
only occur when the rolling rate is much larger than the displacement
rate. 
Figure~\ref{fig:2_particles}b shows that a reduction in the
eventual tangential force is only realized after a long and sustained
cumulative rolling motion: only after the two particles have rolled
across each other a distance of several contact radii ``$a$''. 
Such vigorous and sustained rolling is unlikely to 
occur during brief collisional
encounters between particles, as might occur in the collisions of
a rapidly sheared granular gas. 
The effect of rolling creep-friction
might be more significant, however, during slow, sustained motions,
as would occur in dense granular flows, where contacts persist for
longer periods of gross deformation. 
This possibility is investigated
in a second application of the creep-friction model.

\subsection{Triaxial compression of 4096-particle system}

In the second system, two assemblies of 4096 densely packed spheres
are slowly loaded in triaxial compression. Because the creep-friction
equation~(\ref{eq:dQ3Db}) suggests that the effect of rolling depends
upon the ratio $a/R$, the two assemblies differ in the radii of their
contact areas relative to the particle radius. The first assembly
has relatively small contacts (small ratios $a/R$), as it contains
relatively hard particles that are confined at a lower pressure. The
second assembly contains softer particles and is confined at a higher
pressure, so that the contacts are larger (large ratios $a/R$).
\par
Both assemblies contain the same 4096 spheres, ranging in size from
0.4$D_{50}$ to 1.5$D_{50}$, where $D_{50}$ is the median diameter.
The particles are tightly packed into a cube container having periodic
boundaries on all sides. 
Starting with this same initial arrangement
of spheres, the two assemblies were allowed to adjust to the particular
elastic properties of their particles and their confining stress,
yet producing about the same bulk density 
(see Table~\ref{table:DEM_properties}).
\begin{table}
\caption{Characteristics of two DEM assemblies\label{table:DEM_properties}}
\centering
\begin{tabular}{lcc}
\hline 
 & \multicolumn{2}{c}{Values}\\
\cline{2-3} 
 & Smaller & Larger\\
Characteristic & contacts & contacts\\
\hline 
Particles & \multicolumn{2}{c}{4096}\\
Particle shape & \multicolumn{2}{c}{Spheres}\\
Particle shear modulus, $G$ & 29 GPa & 1.2 GPa\\
Particle Poisson ratio, $\nu$ & 0.15 & 0.34\\
Friction ratio, $\mu$ & 0.50 & 0.50\\
Confining pressure, $p^{\prime}$ & 10 kPa & 1800 kPa\\
Initial indentation, $\zeta/R$ & 0.0054\% & 1\%\\
Initial contact radius, $a/R$ & 0.73\% & 10\%\\
Initial void ratio & 0.513 & 0.461\\
Initial avg. coordination no. & 1.86 & 2.46\\
\hline 
\end{tabular}
\end{table}
Particles in the first assembly are assigned the elastic properties
of quartz, and this assembly was initially confined with a mean stress
of 10kPa (for a sand, this pressure would be attained at a burial
of about 0.5m). Under these conditions, the average contact indentation
$\zeta$ is about 0.0054\% of the mean particle radius $R$, and the
average contact radius is about 0.73\% of the particle radius 
(Table~\ref{table:DEM_properties}).
Particles in the second assembly are given the softer properties of
a polymeric plastic and are confined to the higher mean stress of
1.8MPa, so that the average contact indentation is about 1\% of the
mean particle radius, and the mean contact radius is about 10\% of
the particle radius. 
Following the initial isotropic compaction, a deviatoric loading is
applied in the form of triaxial compression:
compressing the assembly in one direction at a constant rate
of strain while the assembly is allowed to expand equally in the other
two directions so that a constant mean stress is maintained within
the assembly. 
With each assembly, two simulations are run. 
In one simulation, 
the effect of creep-friction is included, by using Eq.~(\ref{eq:dQ3Db})
to compute the tangential contact force. 
With the second simulation,
creep-friction is negated by using a standard Cattaneo-Mindlin model
of tangential force: $d\mathbf{Q}=\mathcal{C}_{\text{C-M}}^{-1}\left(d\overline{\mathbf{u}}^{\text{disp}}\right)$.
In both contact models, the J\"{a}ger algorithm was used to implement
the stiffness function $\mathcal{C}_{\text{C-M}}^{-1}$ 
\cite{Kuhn:2011a}.
\par
\begin{figure}
\centering
\subfigure[Smaller contacts]{\includegraphics[scale=0.92]{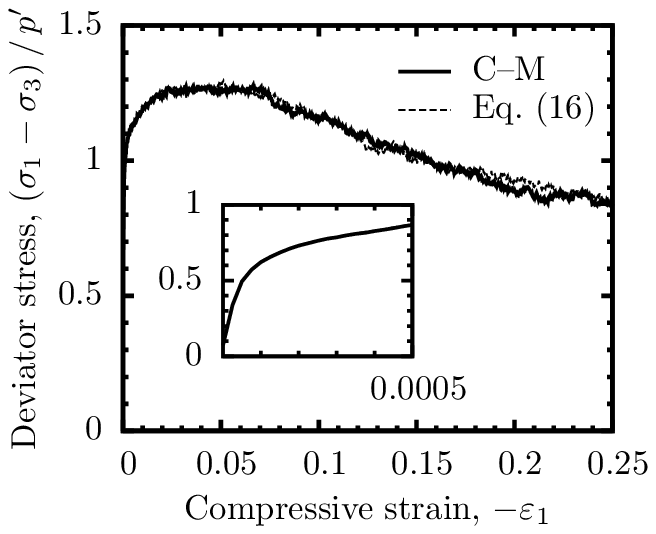}}
\subfigure[Larger contacts]{\includegraphics[scale=0.92]{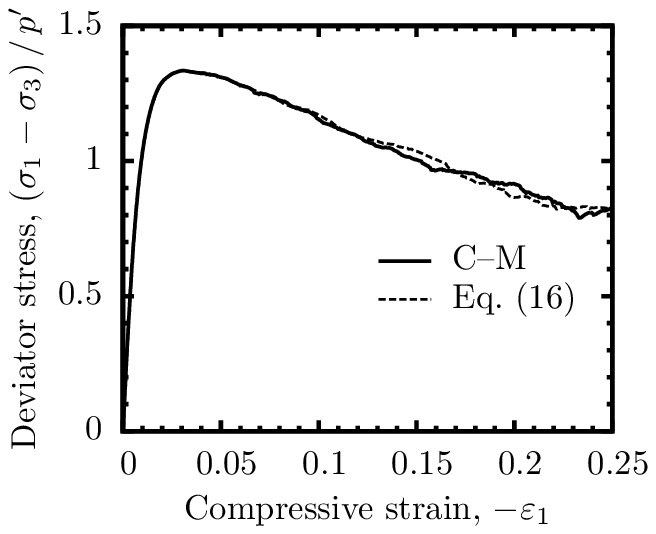}}
\caption{Results of loading two dense assemblies of 4096 spheres: one with
small contacts, the other with large contacts (see Table~\ref{table:DEM_properties}).
Each plot shows results of two contact models: the standard Cattaneo-Mindlin
(C-M) model without creep-friction, and the transient creep-friction
model of Eq.~(\ref{eq:dQ3Db}).\label{fig:4096results}}
\end{figure}

The results of four simulations are shown in Fig.~\ref{fig:4096results}:
for assemblies with small and with large contacts, and for simulations
with and without creep-friction.%
Although different line styles are used for plotting the two contact
models, the results are almost indistinguishable: the inclusion of
creep-friction in the contact model has almost no discernible effect
on the macro-scale behavior of either assembly. 
\par
We now consider the micro-scale behavior in three respects: (1) a
comparison of the relative amounts of rolling for the two assemblies
(assemblies with small and large contacts), (2) a comparison of the
relative amounts of rolling for the contact models with and without
creep-friction, and (3) the effect of creep-friction on the incremental
contact response.
\par
Table~\ref{tab:Results1} compares the relative amounts of rolling
among the contacts of the two assemblies. 
\begin{table}
\caption{Micro-scale results of DEM simulations. 
Distributions of relative rolling are shown for two assemblies, 
each with two contact models.\label{tab:Results1}}
\centering
\begin{tabular}{llcccc}
\hline 
 &  & \multicolumn{4}{c}{$\left|d\mathbf{u}^{\text{roll}}\right|/\left|d\overline{\mathbf{u}}^{\text{disp}}\right|$}\\
\cline{3-6} 
 &  &  & \multicolumn{3}{c}{Quartiles}\\
\cline{4-6} 
Assembly & Contact model & Mean & 1st & Median & 3rd\\
\hline 
Small contacts & Cattaneo-Mindlin & 40 & 3.5 & 18 & 43\\
 & Creep-friction, Eq.~(\ref{eq:dQ3Db}) & 43 & 3.2 & 20 & 49\\
Large contacts & Cattaneo-Mindlin & 8.5 & 1.2 & 3.7 & 8.5\\
 & Creep-friction, Eq.~(\ref{eq:dQ3Db}) & 8.7 & 1.2 & 4.4 & 9.8\\
\hline 
\end{tabular}
\end{table}
The statistics are taken from snapshots at a compressive strain of
23\% (see Fig.~\ref{fig:4096results}). The table gives the ratios
$1/\xi=\left|d\mathbf{u}^{\text{roll}}\right|/\left|d\overline{\mathbf{u}}^{\text{disp}}\right|$
of the rolling rates to the rates of translational displacements among
the $\approx$10,000 contacts at this strain. 
Values are given for
the two assemblies, each with the two contact models: 
the standard Cattaneo-Mindlin
contact model (without a creep-friction rolling effect) and the transient
creep-friction model of Eq.~(\ref{eq:dQ3Db}). Two trends are apparent
in Table~\ref{tab:Results1}:
\begin{itemize}
\item The assembly with the larger contacts (i.e., larger ratio of contact
radius to particle radius, $a/R$, due to its softer particles and
larger confining stress) exhibits less rolling than the assembly with
smaller contacts. That is, the ratios $\left|d\mathbf{u}^{\text{roll}}\right|/\left|d\overline{\mathbf{u}}^{\text{disp}}\right|$
are larger for the assembly with the smaller contacts. This result,
perhaps unexpected, is discussed below.
\item The inclusion of creep-friction is seen to reduce the amount of rolling
at the contacts~--- but only slightly. This small, nearly indistinguishable
change at the micro-scale is consistent with the nearly identical
macro-behavior shown in Fig.~\ref{fig:4096results}.
\end{itemize}
With the first trend, the amount of rolling is shown to depend upon
the relative size of the contacts. This trend is found by comparing
the results from assemblies with small and large contacts, since the
average ratio $a/R$ differs by a factor of more than 13 for these two
assemblies. In one sense, the difference in rolling rates is unexpected,
as neither contact model includes any contact moments that would overtly
inhibit rolling within the assembly having the larger contacts. Moreover,
the inclusion of creep-friction in the contact model has a minimal
effect upon the rolling rates, as noted in the second trend. Although
unexpected, the more vigorous micro-scale rolling for small confining
pressures (i.e., small contacts) 
is likely related to certain macro-scale behaviors in granular
materials. The pressure-dependent strength of granular materials is
a signature characteristic of their mechanical behavior: the strength
of a granular material increases with the confining pressure. Strength
is not always proportional, however, to the confining pressure, an
observation that is usually attributed to the more vigorous dilation
that occurs at low pressures. The dependence of the rolling rates
upon the contact size $a/R$ (and, hence, on the confining pressure)
suggests a type of micro-scale behavior that controls the rate of
dilation. When two particles roll, their motions conform to the curvatures
of the two surfaces (Eq.~\ref{eq:durollgeneral}) and their indentations
(or overlaps) are preserved; but when particles slide, the original
points of contact on the two surfaces are offset. That is, the two
forms of interaction~--- rolling and sliding~--- likely produce
different volumetric responses. 
\par
The relative effect of creep-friction on the micro-scale behavior
is explored with Table~\ref{tab:Results2}, which gives statistics
among the $\approx$10,000 contacts at a strain of 23\%.%
\begin{table}
\caption{Micro-scale results of DEM simulations. 
Distributions of the instantaneous effect of creep-friction 
on the displacement $d\overline{\mathbf{u}}^{\text{disp}}$.
The transient creep-friction model of Eq.~(\ref{eq:dQ3Db}) is used
in both simulations.\label{tab:Results2}}
\centering
\begin{tabular}{lcccc}
\hline 
 & \multicolumn{4}{c}{Relative effect of creep-friction, Eq.~\ref{eq:RelativeEffect}}\\
\cline{2-5} 
 &  & \multicolumn{3}{c}{Quartiles}\\
\cline{3-5} 
Assembly & Mean & 1st & Median & 3rd\\
\hline 
Small contacts & 0.10 & 0.009 & 0.04 & 0.11\\
Large contacts & 0.30 & 0.049 & 0.14 & 0.34\\
\hline 
\end{tabular}
\end{table}
The transient creep-friction model of Eq.~(\ref{eq:dQ3Db}) was
used in assemblies having small and large contacts. In the equation,
the stiffness function $\mathcal{C}_{\text{C-M}}^{-1}$ is applied
to a contact displacement $d\overline{\mathbf{u}}^{\text{disp}}$
that has been altered by the concurrent rolling increment $d\mathbf{u}^{\text{roll}}$.
The relative extent of this alteration, therefore, serves as a measure
of the relative effect of creep-friction on individual contacts. 
Table~\ref{tab:Results2}
presents statistics of the alteration relative to the magnitude of
the displacement itself, in the form of the ratio
\begin{equation}
\left.\frac{8\mu}{(1-\nu)C_{11}}\frac{a}{R}\left(1-\left(1-\frac{|\mathbf{Q}|}{\mu P}\right)^{1/3}\right)\,|d\mathbf{u}^{\text{roll}}|\right/\left|d\overline{\mathbf{u}}^{\text{disp}}\right|\label{eq:RelativeEffect}
\end{equation}
Two trends are apparent in Table~\ref{tab:Results2}:
\begin{itemize}
\item Although rolling proceeds vigorously in both assemblies, the relative
effect of creep-friction is modest. For example, 
the previous Table~\ref{tab:Results1}
shows that the mean rate of rolling is about 40 times the rate of
displacement for the assembly with smaller contacts.
yet the relative effect of creep-friction has a mean value of only
0.10 (Table~\ref{tab:Results2}). 
This small effect is due to a relative
contact size, the ratio $a/R$, 
that is much less than one (Eqs.~\ref{eq:dQ3Db} and~\ref{eq:RelativeEffect}).
\item
Creep-friction has a greater influence on incremental behavior
in the assembly with larger contacts. Even though rolling is less
vigorous among larger contacts, the larger ratio $a/R$ enhances its
influence.
\end{itemize}
Although the incremental effect of creep-friction is shown to be modest
in Table~\ref{tab:Results2}, the influence is certainly not small.
One might expect that creep-friction would 
have a modest effect on the macro-scale
stress-strain behavior, but instead, the influence of creep-friction
is shown to be altogether insignificant in Fig.~\ref{fig:4096results}.
This contradictory result is attributed to three reasons. First, the
effect of rolling on the tangential force will be minimal unless the
relative rate of rolling (i.e., the inverse creepage $1/\xi=\left|d\mathbf{u}^{\text{roll}}\right|/\left|d\overline{\mathbf{u}}^{\text{disp}}\right|$)
is greater than 10 or more, as is shown in Fig.~\ref{fig:2_particles}a.
Second, even when the rolling rate is large, this rolling must be
sustained for long periods
(i.e., rolling distances) in order to significantly reduce the
friction limit (see Fig.~\ref{fig:2_particles}b). The inter-particle
motions within a dense assembly are fairly erratic, in that contact
motions frequently change direction over rather small periods of strain
(e.g., \S1 of \cite{Kuhn:2011a}), 
not allowing the sustained, uniform
motions that are necessary for rolling to have a significant effect
on the contact forces. 
Finally, numerous studies have shown that the
bulk strength of a granular material is insensitive to the 
friction coefficient~$\mu$,
since strength results primarily from an anisotropy of the normal
contact forces 
(e.g.,~\cite{Thornton:2000a}). 
That is, any small
influence of creep-friction upon the contact tangential forces will
have an even smaller effect upon the material's bulk strength.

\section{Conclusions}

The paper identifies two categories of rotational resistance and dissipation
between contacting bodies: one category results in a contact moment,
the other results in micro-slip. 
Until now, only the first category
has been applied in DEM simulations through the use
of rotational springs, dampers, and sliders. 
The latter form of rotational
resistance~--- termed creep-friction~--- is investigated in the paper. 
This mechanism complements the pure translational sliding
of a Cattaneo-Mindlin mechanism, a mechanism that is widely implemented
in DEM codes. 
Although exact solutions are available for pure, steady-state
creep-friction and for pure Cattaneo-Mindlin sliding, DEM simulations
require solution of the more complex, transitional behavior between
these two extremes (an intermediate condition called the ``transient
problem'' in the rolling friction literature). 
An approximate solution
to the three-dimensional transient problem is proposed. 
The solution
is consistent with both extremes of 
pure creep-friction and pure Cattaneo-Mindlin sliding,
and it is also consistent with a two-dimensional solution that has
been verified with elastic analysis.
\par
When applied to a system of two spheres, creep-friction is shown 
to be significant only when the particle motions are dominated by rolling,
when translational sliding is minimal, and when such motions are sustained
for long rolling distances while the particles remain in persistent
contact. 
DEM simulations with sphere assemblies demonstrate that the
creep-friction mechanism has a 
modest effect on the micro-scale behavior but an almost
imperceptible effect on the observed macro-scale behavior. 
These results are reassuring, as the creep-friction
mechanism has been overlooked during the past several decades 
in which DEM has been used for investigating granular behavior.

\section*{Acknowledgements}

This material is based upon work supported by the National Science Foundation  under Grant No. NEESR-936408.




\end{document}